\documentclass[preprint,showpacs,preprintnumbers]{revtex4}
\usepackage{amsmath}
\usepackage{graphicx}
\usepackage{dcolumn}
\usepackage{bm}
\usepackage{epsfig}
\usepackage[T1]{fontenc}
\usepackage{ae,aecompl}
\usepackage{epstopdf, epsfig}
\usepackage{color}
\usepackage{appendix}

\begin{document}

\title{Kinetic modeling of multiphase flow based on simplified Enskog equation}
\author{Yudong Zhang$^{1}$, Aiguo Xu$^{2,3}$\footnote{
Corresponding author. E-mail: Xu\_Aiguo@iapcm.ac.cn}, Jingjiang Qiu$^{1}$, Hongtao Wei$^{1}$, Zung-Hang Wei$^{1}$\footnote{
Corresponding author. E-mail: profwei@zzu.edu.cn}}
\affiliation{1, School of Mechanics and Safety Engineering, Zhengzhou University, Zhengzhou 450001, China\\
2, Laboratory of Computational Physics, Institute of Applied
Physics and Computational Mathematics, P. O. Box 8009-26, Beijing 100088,
China \\
3, Center for Applied Physics and Technology, MOE Key Center for High Energy
Density Physics Simulations, College of Engineering, Peking University,
Beijing 100871, China}
\date{\today }

\begin{abstract}
A new kinetic model for multiphase flow was presented under the framework of the discrete Boltzmann method (DBM). Significantly different from the previous DBM, a bottom-up approach was adopted in this model. The effects of molecular size and repulsion potential were described by the Enskog collision model; the attraction potential was obtained through the mean-field approximation method. The molecular interactions, which result in the non-ideal equation of state and surface tension, were directly introduced as an external force term.
Several typical benchmark problems, including Couette flow, two-phase coexistence curve, the Laplace law, phase separation, and the collision of two droplets, were simulated to verify the model. Especially, for two types of droplet collisions, the strengths of two non-equilibrium effects, $\bar{D}_2^*$ and $\bar{D}_3^*$, defined through the second and third order non-conserved kinetic moments of $(f - f ^{eq})$,
are comparatively investigated, where $f$ ($f^{eq}$) is the (equilibrium) distribution function. It is interesting to find that during the collision process, $\bar{D}_2^*$ is always significantly larger than $\bar{D}_3^*$, $\bar{D}_2^*$ can be used to identify the different stages of the collision process and to distinguish different types of collisions.
 The modeling method can be directly extended to a higher-order model for the case where the non-equilibrium effect is strong, and the linear constitutive law of viscous stress is no longer valid.
\end{abstract}

\pacs{ 05.20.Dd, 51.10.+y, 47.11.-j \\
\textbf{Keywords:} multiphase flow, discrete Boltzmann method, Enskog equation, non-equilibrium characteristics}

\maketitle

\section{Introduction}
Multiphase flow is ubiquitous in nature and industrial processes, such as the formation of raindrops, fabrication of pills, oil and gas exploitation, micro-fluidic chip\cite{Micro-pill,droplet,gas-oil,Microfluidics}, etc. Understanding the hydrodynamic and thermodynamic characteristics of multi-phase flow is of great significance for fundamental research and engineering applications.

Generally, multiphase flow refers to the fluid flow consisting of two or more phases. It is more complicated than a single-phase flow process, due to phase interfaces and the interaction between the different phases \cite{Multiphase-book}. The multiphase flow process is often accompanied by abundant interfacial dynamic behaviors, including interface generation, movement, deformation, fragmentation, and fusion, etc. These complex interfacial behaviors are challenging to measure experimentally and very difficult, if not impossible, to obtain analytic solutions.

As computation power increased in recent decades, numerical simulation has gradually become a critical alternative approach to study multiphase flow problems \cite{Saurel2018,Frezzotti2019,Worner2012}. In general, the numerical simulation methods of multiphase flow can be divide into three levels: macroscopic, microscopic, and mesoscopic methods. The macroscopic approaches are mainly based on the numerical solutions of the hydrodynamic (Navier-Stokes, NS) equations combined with interface capture schemes \cite{Saurel2018}. Although those methods have been successfully applied to the large-scale multiphase flow problems, they fail to describe the micro-nano scale interface structure because the equations based on the continuum assumption may not be applicable in the micro-nano scale \cite{Worner2012,Zhang2019}. In addition, the fluctuations are significant with the decreasing of system size and can even perceivably affect the flow characteristics at the micro-nano scale \cite{MoselerScience2000}. The surface tension and viscosities might need to be further considered with the effect of fluctuation. The microscopic model mainly refers to the molecular dynamic (MD) method, which is more fundamental and accurate \cite{MD-2020}. However, the computational burden of MD simulation is often unaffordable; consequently, the time and space scales it can simulate are limited. As a bridge between macroscopic and microscopic, the mesoscopic model has developed rapidly over the past three decades \cite{Wolfram,Chen-LBM1998,Succi,He2002,Qin2006}. It has been successfully applied in a wide variety of multiphase problems, including wetting, droplet collision and splashing, boiling and evaporation, phase separation, and hydrodynamic instability \cite{Qin2015,Li2016,Timmbook,Saurel2018}, etc.

The mesoscopic model for multiphase flow, also known as the kinetic model, is mainly based on the Boltzmann equation. The vast majority, if not all, of the mesoscopic multiphase flow models in the literature, are developed from the lattice Boltzmann model (LBM), including the color-gradient model \cite{Color1993}, the Shan-Chen model \cite{Shan-chen1993}, the free energy model \cite{free-energy1995,XGL-PRE2003}, and the phase-field model \cite{He-chen-zhang1999,Liang-PoF2019,Wang2019,Sun-AML2019}, etc. It is undeniable that they are very convenient in modeling multiphase flow processes, and have achieved great success in the research of various multiphase flow problems \cite{Li2016}. However, it should be noted that the standard LBM, on which most multiphase flow models base, is just a solver of the incompressible Navier-Stokes (NS) equations. Although it may be more convenient to deal with the interparticle forces and the complicated boundaries \cite{Chen-LBM1998}, it can not provide kinetic information beyond the NS equations.

Based on the thermal finite different LBM of Watari and Tsutahara \cite{Watari2003}, Gonnella et al. first proposed the thermal multiphase model \cite{Gonnella2007}, in which the nonideal gases and the surface tension effects are introduced by an extra term $I_{ki}$. This multiphase flow model can correctly reproduce the transport equations of nonideal fluid established by Onuki \cite{Onuki2005}, through a Chapman-Enskog multiscale expansion. Gan et al.\cite{GanFFT-2012} further developed a thermal multiphase model with negligible spurious velocities by introducing the Windowed Fast Fourier Transform scheme when calculating the spatial derivatives in the convection term and the external force term. Then morphological characteristics of phase separation in the thermal systems were systematically studied \cite{Gan-morphological2012}.


In the early studies, the non-equilibrium behaviors in various complex flow were all those described by hydrodynamic equations. For the convenience of description, we refer those behaviors to as Hydrodynamic Non-Equilibrium (HNE) behaviors.
Besides the HNE, the most relevant Thermodynamic Non-Equilibrium (TNE) practices in
various complex flow systems are attracting more and more attention with time\cite{XuA2012,xu2015aps,xu2016,DBM-book2018}.

It is understandable that when the TNE is very weak, the loss of considering TNE is not meaningful.
But when the TNE is strong, the situation will be significantly different.
For example, it has been shown that TNE's existence directly affects the density, temperature, and pressure,
as well as the magnitude and direction of flow velocity. Without considering TNE, the density, flow velocity,
temperature, and pressure given will have a significant deviation \cite{Lin-2019-CNF}.
The existence of TNE is the underlying cause of the appearance of heat flow and viscous stress. If insufficiently considered
(considered only the linear response part), the amplitude of heat flow and viscous stress obtained may
be too large\cite{Gan2018}.

When the strength of TNE beyond some threshold value,
it may change the directions of heat flow and viscous stress.
If the TNE is not sufficiently considered (considered only the linear response part),
the obtained viscous stress and heat flow maybe, even in the wrong directions \cite{Zhang2019}.
In addition, in the phase separation system, both the mean TNE strength\cite{gan2015}
and the entropy production rate, one of the quantities describing the TNE effects\cite{SoftMatter2019},  increase with time
in the spinnodal decomposition stage and decrease with time in the domain growth stage. Therefore, both the peak value points of the mean TNE strength and the
entropy production rate can work as physical criteria to discriminate the two stages\cite{gan2015,xu2016,SoftMatter2019}.
In the system with combustion, the TNE behaviors help to better understand the
physical structures of the von Neumann peak and various non-equilibrium detonation\cite{xu2015aps,yan2013,xu2015pre,Lin2016CNF,zhang2016CNF,Lin-2018-CAF,Lin-2017-SR,Xu-2018-FoP,Lin-2019-CNF}.
In the system with Rayleigh-Taylor instability, the TNE behaviors around interfaces have been used to physically identify and distinguish various interfaces and design relevant
interface-tracking schemes. With increasing the compressibility, more observable TNE kinetic modes appear for given observation precision\cite{Lai2016,DBM-book2018}. The correlation between the mean density nonuniformity and mean TNE strength is almost 1.
The correlation between the mean temperature nonuniformity and mean Non-Organized Energy Flux (NOEF) is almost 1. The correlation between the mean flow velocity nonuniformity and
the mean Non-Organized Momentum Flux (NOMF) is also high, but generally less than 1 \cite{Chen2016,DBM-book2018}. The TNE effect helps to understand the effect of system dispersion
(described by Knudsen number) on its kinetic behaviors\cite{Ye2020}.
In the system with Kelvin-Helmholtz instability, via some defined TNE quantity, for example, the heat flux intensity, the density interface and temperature interface can both be observed so that the material mixing and energy mixing
in the Kelvin-Helmholtz instability evolution can be investigated simultaneously\cite{Zhang2019,Gan2019}. The TNE behaviors were used to understand the mixing entropy in multi-component flows better\cite{Lin2017PRE}.

Since the traditional hydrodynamic model is incapable of capturing enough TNE behaviors,
the works mentioned above on both HNE and TNE resorted to the recently proposed discrete Boltzmann method (DBM)\cite{XuA2012,DBM-book2018}. Compared with the standard LBM, the model system's evolution does not resort to the simple ``propagation + collision" scenario of virtual particle which is inherited from the lattice-gas-automaton model. Any suitable numerical scheme can be used to solve the discrete Boltzmann equation(DBE) according to the specific case. The non-conserved kinetic moments of $(f_i - f^{eq}_i)$ are used to describe how the system deviates from its thermodynamic equilibrium state and to fetch information on the TNE behaviors, where $f_i$ ($f^{eq}_i$) is the discrete (equilibrium) distribution function and $i$ is the index of discrete velocity.  The kinetic moment relations to be based in the construction process of DBM are dependent on the research perspective and the non-equilibrium extent which the DBM aims to describe\cite{DBM-book2018}.
Besides by theory, results of DBM have been confirmed and supplemented by results of molecular dynamics\cite{kw2016,kw2017,kw2017b}, direct simulation Monte Carlo\cite{Zhang2019,Meng2013JFM} and experiment\cite{Lin-2017-SR}.


The existing multiphase flow models based on DBM is a semi-kinetic and semi-top-down approach. The discrete Boltzmann equation is modified by an extra term to be consistent with the correct hydrodynamic equations of nonideal fluid. The specific forms of the coefficients (expressed as the gradient of macroscopic quantities) in the external force term are determined by the Chapman-Enskog multiscale expansion \cite{Gonnella2007,GanSoftmatter2016}. Therefore, in the modeling of multiphase flow DBM, the hydrodynamic equations of nonideal fluid need to be known first. This partly limits the ability of DBM to study the non-equilibrium complex flow, because the hydrodynamic equations are often unknown when the non-equilibrium strength is strong. In this paper, this restriction is addressed by developing a bottom-up kinetic modeling method of multiphase flow under the framework of DBM.

The rest of this paper is organized as follows. In Sec. \ref{sec2}, the simplified Enskog equation is derived, and the multiphase model based on the discrete Enskog equation is presented. Several benchmark tests of multiphase flow are simulated to verify the new model, and the non-equilibrium characteristics during the collision of two droplets are analyzed in Sec.\ref{sec3}. Section \ref{sec4} concludes the present paper.

\section{Models and methods}\label{sec2}
\subsection{Enskog equation and its simplification}

According to molecular kinetic theory, the evolution equation of molecular velocity distribution function reads
\begin{equation}\label{Evolution-equation1}
\frac{{\partial f}}{{\partial t}} + {\bf{v}} \cdot \nabla f + {\bf{a}} \cdot {\nabla _{\bf{v}}}f = {\left( {\frac{{\partial f}}{{\partial t}}} \right)_c} ,
\end{equation}
where $f=f(\textbf{r},\textbf{v},t)$ is the molecular velocity distribution function, $\mathbf{r}$ and $\mathbf{v}$ represent the position space and velocity space coordinates, respectively, $\mathbf{a}$ is the acceleration generated by the total extra force. ${\left( {\frac{{\partial f}}{{\partial t}}} \right)_c}$ denotes the rate of change in distribution function due to the collisions between molecules. Given that the probability of two molecules colliding is much higher than that of three or more molecules colliding simultaneously, the hypothesis of two-body collision is reasonable.

Based on the elastic molecular collision model, when the molecular volume effect is ignored, the collision term in the Boltzmann equation can be obtained as \cite{Shenqing-book}
\begin{equation}\label{Eq-Collisonterm1}
{\left( {\frac{{\partial f}}{{\partial t}}} \right)_c} = \int_{ - \infty }^\infty  {\int_0^{4\pi } {\left( {{f^*}f_1^* - f{f_1}} \right){v_r}\sigma d\Omega d{{\bf{v}}_1}} },
\end{equation}
where $f^*$ and $f^*_1$ represent the post-collision molecular velocity distribution function with velocity $\mathbf{v}^*$ and $\mathbf{v}^*_1$, respectively, $f$ and $f_1$ represent the pre-collision molecular velocity distribution function with velocity $\mathbf{v}$ and $\mathbf{v}_1$, respectively. ${v_r} = \left| {{\bf{v}} - {{\bf{v}}_1}} \right|$ is the value of the relative velocity, which remains unchanged pre to and after the collision. $\sigma$ and $\Omega$ denote the differential collision cross-section and solid angle, respectively.

However, this assumption is not appropriate for dense gases or liquid. With increasing the number density, compared with the mean distance between neighboring molecules, the size of a gas molecule is no longer negligible anymore, the volume effects of the molecule should be taken into account, the Boltzmann collision operator should be replaced by the Enskog collision operator which reads \cite{Chapman-book}
\begin{equation}\label{Eq-Collisonterm2}
\begin{array}{l}
{\left( {\frac{{\partial f}}{{\partial t}}} \right)_E} = \int_{ - \infty }^\infty  {\int_0^{4\pi } {\left[ {\chi \left( {{\bf{r}} + \frac{{{d_0}}}{2}{{{\bf{\hat e}}}_r}} \right){f^*}\left( {\bf{r}} \right)f_1^*\left( {{\bf{r}} + {d_0}{{{\bf{\hat e}}}_r}} \right)} \right.} } \\
\left. {  {\kern 86pt}  - \chi \left( {{\bf{r}} - \frac{{{d_0}}}{2}{{{\bf{\hat e}}}_r}} \right)f\left( {\bf{r}} \right){f_1}\left( {{\bf{r}} - {d_0}{{{\bf{\hat e}}}_r}} \right)} \right]{v_r}\sigma d\Omega d{{\bf{v}}_1},
\end{array}
\end{equation}
where $d_0$ is the diameter of the hard-sphere molecules and $\chi$ represents the collision probability correction considering the molecular volume effect. By using the Taylor expansion and keeping to the first derivative term, one can get that
\begin{equation}\nonumber \label{Eq-Tayloar1}
\chi ( {{\bf{r}} + \frac{{{d_0}}}{2}{{{\bf{\hat e}}}_r}}) = \chi \left( {\bf{r}} \right) + \frac{{{d_0}}}{2} \nabla \chi \cdot {{\bf{\hat e}}_r}
\end{equation}

\begin{equation}\nonumber \label{Eq-Tayloar2}
\chi ( {{\bf{r}} - \frac{{{d_0}}}{2}{{{\bf{\hat e}}}_r}} ) = \chi \left( {\bf{r}} \right) - \frac{{{d_0}}}{2} \nabla \chi \cdot {{\bf{\hat e}}_r}
\end{equation}

\begin{equation}\nonumber \label{Eq-Tayloar3}
f_1^*\left( {{\bf{r}} + {d_0}{{{\bf{\hat e}}}_r}} \right) = f_1^*\left( {\bf{r}} \right) +
{d_0} \nabla f_1^* \cdot {{\bf{\hat e}}_r}
\end{equation}

\begin{equation}\nonumber \label{Eq-Tayloar4}
f_1^{}\left( {{\bf{r}} - {d_0}{{{\bf{\hat e}}}_r}} \right) = f_1^{}\left( {\bf{r}} \right)
-{d_0} \nabla f_1 \cdot {{\bf{\hat e}}_r}
\end{equation}
Consequently, the collision term in Eq. (\ref{Eq-Collisonterm2}) becomes
\begin{equation}\label{Eq-Collisonterm3}
\begin{array}{l}
{\left( {\frac{{\partial f}}{{\partial t}}} \right)_E} = \chi \int_{ - \infty }^\infty  {\int_0^{4\pi } {\left( {{f^*}f_1^* - f{f_1}} \right){v_r}\sigma d\Omega d{{\bf{v}}_1}} } \\
{\kern 16pt} + {d_0}\chi \int_{ - \infty }^\infty  {\int_0^{4\pi } {\left( {{f^*}\nabla f_1^* + f\nabla {f_1}} \right) \cdot {{{\bf{\hat e}}}_r}{v_r}\sigma d\Omega d{{\bf{v}}_1}} } \\
{\kern 16pt}  + \frac{1}{2}{d_0}\int_{ - \infty }^\infty  {\int_0^{4\pi } {\nabla \chi  \cdot {{{\bf{\hat e}}}_r}\left( {{f^*}f_1^* - f{f_1}} \right){v_r}\sigma d\Omega d{{\bf{v}}_1}} }
\end{array}
\end{equation}
where $\chi$, ${f^*_1}$, and ${f_1}$ are all at $\mathbf{r}$.
If $f$ in the latter two terms are approximated by the local equilibrium distribution function $f^{eq}$,
\begin{equation}\label{Eq-feq}
{f^{eq}} = \rho \frac{1}{{{{\left( {2\pi RT} \right)}^{3/2}}}}\exp [ { - \frac{{{{({\mathbf{v}} - {\mathbf{u}})}^2}}}{{RT}}} ].
\end{equation}
Then the Enskog collision operator becomes
\begin{equation}\label{Eq-Collisonterm4}
\begin{array}{l}
{\left( {\frac{{\partial f}}{{\partial t}}} \right)_E} = \chi \int_{ - \infty }^\infty  {\int_0^{4\pi } {\left( {{f^*}f_1^* - f{f_1}} \right){v_r}\sigma d\Omega d{{\bf{v}}_1}} } \\
- {f^{eq}}b\rho \chi \left\{ {\left( {{\mathbf{v}} - {\mathbf{u}}} \right) \cdot \left[ {\frac{2}{\rho }
\nabla \rho + \frac{3}{5T} \nabla T \left( {\frac{{{{\left( {{\bf{v}} - {\bf{u}}} \right)}^2}}}{{ 2 RT}} - \frac{5}{2}} \right)} \right]} \right.\\
\left. { + \frac{3}{{5RT}}\left( {{\mathbf{v}} - {\mathbf{u}}} \right)\left( {{\mathbf{v}} - {\mathbf{u}}} \right): \nabla \mathbf{u} + \frac{3}{5} \left( { \frac{{{{\left( {{\bf{v}} - {\bf{u}}} \right)}^2}}}{{2RT}}} - \frac{5}{2} \right) } \nabla \cdot \mathbf{u} \right\} - {f^{eq}}b\rho \left( {{\mathbf{v}} - {\mathbf{u}}} \right) \cdot \nabla \chi  ,
\end{array}
\end{equation}
where $b\rho=\frac{2}{3}\pi n {d_0}^3$.

For isothermal systems, the collision term can be reduced to
\begin{equation}\label{Eq-Collisonterm5}
\begin{array}{l}
{\left( {\frac{{\partial f}}{{\partial t}}} \right)_E} = \chi \int_{ - \infty }^\infty  {\int_0^{4\pi } {\left( {{f^*}f_1^* - f{f_1}} \right){v_r}\sigma d\Omega d{{\bf{v}}_1}} } \\
- {f^{eq}}b\rho \chi \left\{ {\left( {{\mathbf{v}} - {\mathbf{u}}} \right) \cdot  {\frac{2}{\rho }
\nabla \rho  } } \right.
\left. { + \frac{3}{{5RT}}\left( {{\mathbf{v}} - {\mathbf{u}}} \right)\left( {{\mathbf{v}} - {\mathbf{u}}} \right): \nabla \mathbf{u} + \frac{3}{5} \left[ { \frac{{{{\left( {{\bf{v}} - {\bf{u}}} \right)}^2}}}{{2RT}}} - \frac{5}{2} \right] } \nabla \cdot \mathbf{u} \right\}  \\
- {f^{eq}}b\rho \left( {{\mathbf{v}} - {\mathbf{u}}} \right) \cdot \nabla \chi   .
\end{array}
\end{equation}
In addition,  the last two terms in the second row of Eq. (\ref{Eq-Collisonterm5}) satisfy the following relationships:
\begin{equation}\label{Eq-collisionM0}
\int {{f^{eq}}b\rho \chi \left\{ {\frac{3}{{5RT}}({\bf{v}} - {\bf{u}})({\bf{v}} - {\bf{u}}):\nabla {\bf{u}} + \frac{3}{5} \left[ { \frac{{{{\left( {{\bf{v}} - {\bf{u}}} \right)}^2}}}{{2RT}}} - \frac{5}{2} \right] \nabla  \cdot {\bf{u}}} \right\}d{\bf{v}}}  = 0  ,
\end{equation}
\begin{equation}\label{Eq-collisionM0}
\int {{f^{eq}}b\rho \chi \left\{ {\frac{3}{{5RT}}({\bf{v}} - {\bf{u}})({\bf{v}} - {\bf{u}}):\nabla {\bf{u}} + \frac{3}{5} \left[ { \frac{{{{\left( {{\bf{v}} - {\bf{u}}} \right)}^2}}}{{2RT}}} - \frac{5}{2} \right] \nabla  \cdot {\bf{u}}} \right\}{\bf{v}}d{\bf{v}}}  = 0  .
\end{equation}
The sum of these two terms have no effect on the evolution of density and momentum \cite{Guo-book}, so they can be ignored in the modeling framework of this paper. As a result, the Enskog collision term can be further simplified as
\begin{equation}\label{Eq-Collisonterm6}
\begin{array}{l}
{\left( {\frac{{\partial f}}{{\partial t}}} \right)_E} = \chi \int_{ - \infty }^\infty  {\int_0^{4\pi } {\left( {{f^*}f_1^* - f{f_1}} \right){v_r}\sigma d\Omega d{{\bf{v}}_1}} } \\
 {\kern 31pt}  - {f^{eq}}b\rho \chi \left( {\mathbf{v} - \mathbf{u}} \right) \cdot \frac{2}{\rho }
 \nabla \rho - {f^{eq}}b\rho \left( {\mathbf{v} - \mathbf{u}} \right) \cdot \nabla \chi  .
\end{array}
\end{equation}

Similar to the modeling of DBM, the first term on the right-hand side can be replaced by the BGK collision operator, consequently Eq. (\ref{Eq-Collisonterm5}) becomes
\begin{equation}\label{Eq-Collisonterm7}
{\left( {\frac{{\partial f}}{{\partial t}}} \right)_E} =  - \frac{1}{\tau }\left( {f - {f^{eq}}} \right) - {f^{eq}}b\rho \chi \left( {{\mathbf{v}} - {\mathbf{u}}} \right) \cdot \nabla \ln \left( {{\rho ^2}\chi } \right)  .
\end{equation}
Now we can get the simplified Enskog model for the approximately incompressible fluid systems when the temperature is uniform, which reads
\begin{equation}\label{Eq-Enskog}
\frac{{\partial f}}{{\partial t}} + {\bf{v}} \cdot \nabla f + {\bf{a}} \cdot {\nabla _{\bf{v}}}f =  - \frac{1}{\tau }\left( {f - {f^{eq}}} \right) - {f^{eq}}b\rho \chi \left( {{\mathbf{v}} - {\mathbf{u}}} \right) \cdot \nabla \ln \left( {{\rho ^2}\chi } \right)  .
\end{equation}

Similar to previous studies, in this work we consider the case where
the force term ${\bf{a}} \cdot {\nabla _{\bf{v}}}f$ can be approximated by
\begin{equation}\label{Eq-Enskog2}
{\bf{a}} \cdot {\nabla _{\bf{v}}}f = {\bf{a}} \cdot {\nabla _{\bf{v}}}f^{eq} =- \frac{{{\mathbf{F}}\cdot( {{\mathbf{v}} - {\mathbf{u}}} )}}{{\rho RT}}{f^{eq}} ,
\end{equation}
where $m$ is the mass of the molecule. As a result, the simplified Enskog equation becomes
\begin{equation}\label{Eq-Enskog3}
\frac{{\partial f}}{{\partial t}} + {\bf{v}} \cdot \nabla f - \frac{{{\mathbf{F}}\cdot( {{\mathbf{v}} - {\mathbf{u}}} )}}{{\rho RT}}{f^{eq}} =  - \frac{1}{\tau }\left( {f - {f^{eq}}} \right) - \frac{{( {{\mathbf{v}} - {\mathbf{u}}} ) \cdot \nabla (b \rho^2 RT \chi)}}{{\rho RT}}{f^{eq}} .
\end{equation}
The last term on the right side of the equation represents the repulsion interaction of molecules.

\subsection{Multiphase flow model based on Enskog equation}

Now the repulsion interaction of molecules is introduced by using the Enskog collision operator instead of the Boltzmann collision operator, then the key to modeling multiphase flow is to incorporate the intermolecular attraction. According to the average filed approximation, the mutual attraction between molecules can be regarded as an average force potential \cite{He-chen-zhang1999}, which reads
\begin{equation}\label{Eq-force-potential1}
\Phi ({{\bf{r}}_1}) = \int_{{r_{12}} > {d_0}} {{\phi _{attr}}({r_{12}})\rho ({{\bf{r}}_2})d{{\bf{r}}_2}}  ,
\end{equation}
where ${r_{12}} = \left| {{{\bf{r}}_1} - {{\bf{r}}_2}} \right|$ is the distance between $\bf{r}_1$ and $\bf{r}_2$, ${\phi _{attr}}({r_{12}})$ denotes the attraction potential.
Expanding $\rho ({{\bf{r}}_2})$ at $\bf{r}_1$ and ignoring terms higher than the second order, $\Phi ({{\bf{r}}_1})$ can be approximated by
\begin{equation}\label{Eq-force-potential2}
\Phi ({{\bf{r}}_1}) =  - 2a{\rho} - K {\nabla ^2}\rho  ,
\end{equation}
where the first term mainly affects the equation of state (EOS) with $a =  - \frac{1}{2}\int_{r > {d_0}} {{\phi _{attr}}\left( {\bf{r}} \right)d{\bf{r}}}$ and the second term contributes to the surface tension. The coefficient of surface tension is $K =  - \frac{1}{6}\int_{r > {d_0}} {{r^2}{\phi _{attr}}\left( {\bf{r}} \right)d{\bf{r}}}$, and it is assumed to be a constant in this work.

Based on the expression of $\Phi ({{\bf{r}}})$ in Eq. (\ref{Eq-force-potential2}), the total external force of molecules at $\bf{r}$ can be calculated as
\begin{equation}\label{Eq-Force-term}
{\mathbf{F}} =  - \rho \nabla \Phi  = \nabla ( a{\rho}^2) + \rho \nabla ( K {{\nabla ^2}\rho }).
\end{equation}
Substituting the expression of ${\mathbf{F}}$ into Eq. (\ref{Eq-Enskog3}) gives
\begin{equation}\label{Eq-Enskog4}
\frac{{\partial f}}{{\partial t}} + {\bf{v}} \cdot \nabla f - \frac{{{\mathbf{F}'}\cdot( {{\mathbf{v}} - {\mathbf{u}}} )}}{{\rho RT}}{f^{eq}} =  - \frac{1}{\tau }\left( {f - {f^{eq}}} \right)  ,
\end{equation}
where ${\mathbf{F}'}=\left[ { - \nabla \psi  + \rho \nabla \left( {K{\nabla ^2}\rho } \right)} \right]$ with $\psi  = b{\rho ^2}RT\chi  - a{\rho ^2}$. There are two terms in ${\mathbf{F}'}$, the first term is related to the EOS and the second one corresponds to the surface tension.

According to Chapman-Enskog multi-scale expansion, the hydrodynamic equations for multiphase flow can be obtained from Eq. (\ref{Eq-Enskog4}) as follows:
\begin{equation}\label{Eq-NS1}
\frac{{\partial \rho }}{{\partial t}} + \nabla \cdot (\rho \mathbf{u}) = 0 ,
\end{equation}
\begin{equation}\label{Eq-NS2}
 \frac{{\partial (\rho \mathbf{u}})}{{\partial t}} + \nabla \cdot (\rho \mathbf{u} \mathbf{u} +P \mathbf{I}+\mathbf{\Pi}) - \rho \nabla (K {\nabla ^2}\rho) = 0 ,
\end{equation}
where $P$ represents the EOS of real gas in the form of
\begin{equation}\label{Eq-Eos}
P = \rho RT\left( {1 + b\rho \chi } \right) - a{\rho ^2}.
\end{equation}
When $\chi  = \frac{{1 - b\rho /8}}{{{{\left( {1 - b\rho /4} \right)}^3}}}$, the Carnahan-Starling EOS can be obtained as
\begin{equation}\label{Eq-EosCS}
P^c = \rho RT\frac{{1 + \eta  + {\eta ^2} - {\eta ^3}}}{{{{(1 - \eta )}^3}}} - a{\rho ^2},
\end{equation}
with $\eta  = \frac{{b\rho }}{4}$. The van der Waals (VDW) EOS
\begin{equation}\label{Eq-EosVDW}
P^v = \frac{\rho RT}{{1 - b\rho }} - a{\rho ^2},
\end{equation}
can also be derived from Eq. (\ref{Eq-Eos}) when $\chi  = \frac{1}{{1 - b\rho }}$.
The $\mathbf{\Pi}$ in Eq. (\ref{Eq-NS2}) represents the viscous stress which has an expression as
\begin{equation}\label{Eq-viscousstress}
{\mathbf{\Pi}} = \mu \left( \nabla \mathbf{u} + \nabla ^{T} \mathbf{u} - \frac{2}{D} \nabla \cdot \mathbf{u}  \right) ,
\end{equation}
where $D$ denotes spatial dimension and $\mu$ is the coefficient of dynamic viscosity, $\mu=\tau \rho R T$. The last term, $ \rho \nabla (K {\nabla ^2}\rho)$, on the left-hand side of Eq. (\ref{Eq-NS2}) denotes the surface tension $\mathbf{F}_{\rm{s}}$, which is equivalent to the following expression \cite{Timmbook}
\begin{equation}\label{Eq-surfacetension2}
{\mathbf{F}_{\rm{s}}} = \nabla \left[ {(  \frac{K}{2}\nabla \rho \cdot \nabla \rho + K\rho {\nabla ^2}\rho ) \mathbf{I} - K \nabla \rho \nabla \rho} \right].
\end{equation}
\subsection{Discrete Enksog model for multiphase flow}
\begin{figure}
	\centering
	\includegraphics[width=0.5\textwidth]{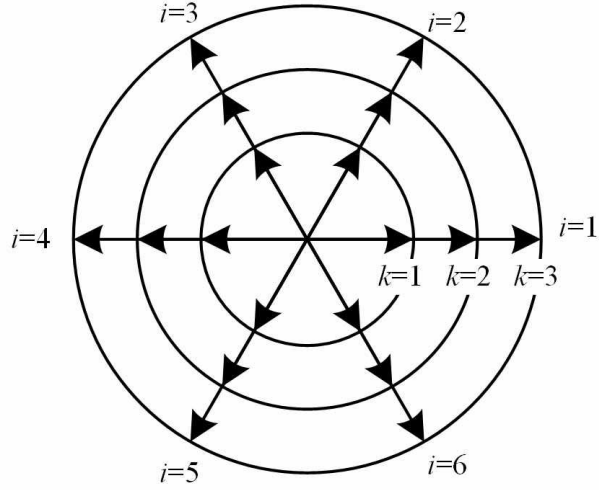}\\
	\caption{Schematic diagram of discrete velocity model.}\label{fig1}
\end{figure}

After the previous derivation, the kinetic equation of multiphase flow is obtained as Eq. (\ref{Eq-Enskog4}). Correspondingly, the evolution of discrete Enskog equation for multiphase flow reads
\begin{equation}\label{Eq-DEE1}
\frac{{\partial {f_{ki}}}}{{\partial t}} + {\mathbf{v}_{ki}} \cdot \nabla f_{ki} - \frac{{({\mathbf{v}_{ki}} - {\mathbf{u}}) \cdot {{\mathbf{F}'}}}}{{\rho RT}}f_{ki}^{eq} =  - \frac{1}{\tau }\left( {{f_{ki}} - f_{ki}^{eq}} \right)  ,
\end{equation}
where $f_{ki} = f_{ki}(\textbf{r},t)$ is the distribution function of the discrete velocity $\mathbf{v}_{ki}$ and the subscript ``$ki$'' indicates the index of discrete velocity. $f_{ki}^{eq}$ is the discrete local equilibrium distribution function; its general expression of any order has been given in the previous literature \cite{Zhang2019}. In the previous Chapman-Enskog multi-scale expansion precess, only the 0th to 3rd kinetic moments are needed to recover Eqs. (\ref{Eq-NS1}) and (\ref{Eq-NS2}). Therefore, for computational efficiency, the discrete local equilibrium distribution function is approximated by the 3rd-order Hermite polynomial which reads
\begin{equation}\label{Eq-fkieq1}
f_{ki}^{eq} = \rho {F_k}\left[ {(1 - \frac{{{u^2}}}{{2T}}) + \frac{1}{{1!}}(1 - \frac{{{u^2}}}{{2T}})\frac{{{\mathbf{v}_{ki}} \cdot {\mathbf{u}}}}{T} + \frac{1}{{2!}}\frac{{{{({\mathbf{v}_{ki}} \cdot {\mathbf{u}})}^2}}}{{{T^2}}} + \frac{1}{{3!}}\frac{{{{({\mathbf{v}_{ki}} \cdot {\mathbf{u}})}^3}}}{{{T^3}}}} \right],
\end{equation}
where $F_k$ is the weight coefficient which can be expressed as
\begin{equation}\label{Eq-fk1}
{F_1} = \frac{{24{T^3} - 4\left( {c_2^2 + c_3^2} \right){T^2} + c_2^2c_3^2T}}{{c_1^2\left( {c_1^2 - c_2^2} \right)\left( {c_1^2 - c_3^2} \right)}},
\end{equation}
\begin{equation}\label{Eq-fk2}
{F_2} = \frac{{24{T^3} - 4\left( {c_1^2 + c_3^2} \right){T^2} + c_1^2c_3^2T}}{{3c_2^2\left( {c_2^2 - c_1^2} \right)\left( {c_2^2 - c_3^2} \right)}},
\end{equation}
\begin{equation}\label{Eq-fk3}
{F_3} = \frac{{24{T^3} - 4\left( {c_2^2 + c_1^2} \right){T^2} + c_2^2c_1^2T}}{{3c_3^2\left( {c_3^2 - c_1^2} \right)\left( {c_3^2 - c_2^2} \right)}},
\end{equation}
and
\begin{equation}\label{Eq-fk0}
{F_0} = 1 - 6\left( {{F_1} + {F_2} + {F_3}} \right).
\end{equation}
where $c_1$, $c_2$, and $c_3$ are the values of three groups of discrete velocities. The scheme of the discrete velocity model is shown in Fig. \ref{fig1}. The value of $c_k$ does not affect the final results as long as the calculation can be performed stably. It should be noted that the new model can be easily extended to higher-order cases only if the discrete local equilibrium distribution function satisfies more kinetic moment relations. Higher-order terms need to be kept in the Hermite polynomials of Eq. (\ref{Eq-fkieq1}) to achieve this aim. Accordingly, more discrete velocities are required to satisfy the higher-order isotropy of the discrete model. More details of the higher-order model are available in Ref. \cite{Zhang2019}.

Once the weight coefficients $F_k$ are calculated, the discrete local equilibrium distribution function $f^{eq}_{ki}$ in Eq. (\ref{Eq-fkieq1}) is also known, then the discrete distribution function in $t+\bigtriangleup t$ (  $f_{ki}^{t+\bigtriangleup t}$) can be solved from $f_{ki}^{t}$ by the following evolution equation
\begin{equation}\label{Eq-DEEevulution2}
f_{ki}^{t + \Delta t} = \frac{1}{\tau }\left( {f_{ki}^t - f_{ki}^{eq}} \right) - {\mathbf{v}_{ki }} \cdot \nabla {f_{ki}^t} + \frac{{({\mathbf{v}_{ki}} - \mathbf{u}) \cdot {{\mathbf{F}'}}}}{{\rho RT}}f_{ki}^{eq}.
\end{equation}
It involves the calculation of spatial derivatives in both the convection term $\nabla {f_{ki}^t}$ and the external force term ${\mathbf{F}'}$. Various types of different schemes, such as finite-difference, Nine-Point Stencils (NPS), Nonoscillatory Non-free-parameter and Dissipative (NND), and Fast Fourier Transform (FFT) schemes, etc. can be used depending on the specific problems.

In addition, from Eq. (\ref{Eq-DEEevulution2}) we see that $f_{ki}$ and its corresponding local equilibrium distribution function $f_{ki}^{eq}$ are both known at a certain time. Consequently, the non-equilibrium quantities $\mathbf{\Delta} _{2}^{*}$ and $\mathbf{\Delta} _{3}^{*}$ can be calculated as
\begin{equation}\label{Eq-Delta2xing}
\mathbf{\Delta}_{2}^* = \sum\limits_{ki} {(f_{ki} - f_{ki}^{eq})} ({\mathbf{v}_{ki}} - {\mathbf{u}})({\mathbf{v}_{ki}} - {\mathbf{u}}),
\end{equation}
\begin{equation}\label{Eq-Delta3xing}
\mathbf{\Delta}_{3}^* = \sum\limits_{ki} {(f_{ki} - f_{ki}^{eq})} ({\mathbf{v}_{ki}} - {\mathbf{u}})({\mathbf{v}_{ki}} - {\mathbf{u}})({\mathbf{v}_{ki}} - {\mathbf{u}}).
\end{equation}
Correspondingly, the quantities of non-equilibrium strength, $D_2^{*}$ and $D_3^{*}$, can be obtained, which are defined as
\begin{equation}\label{Eq-D2xing}
D_2^* = \sqrt {{{\left| {\Delta _{2,xx}^*} \right|}^2} + {{\left| {\Delta _{2,xy}^*} \right|}^2} + {{\left| {\Delta _{2,yy}^*} \right|}^2}}
\end{equation}
and
\begin{equation}\label{Eq-D3xing}
D_3^* = \sqrt {{{\left| {\Delta _{3,xxx}^*} \right|}^2} + {{\left| {\Delta _{3,xxy}^*} \right|}^2} + {{\left| {\Delta _{3,xyy}^*} \right|}^2} + {{\left| {\Delta _{3,yyy}^*} \right|}^2}} ,
\end{equation}
respectively. In previous studies \cite{Zhang2019}, it has been found that some quantities of non-equilibrium strength may perform better than individual non-equilibrium components in describing some characteristics of fluid interfaces. In this work, the role of these non-equilibrium quantities in multiphase flow will be further investigated.

\section{Simulation and discussion}\label{sec3}
\subsection{Couette flow}
As the first test, the Couette flow is simulated to examine the viscous stress calculated by the discrete model with the discrete velocity model shown in Fig. \ref{fig1}. Two parallel plates, filled with fluid between them, are placed in the $y$ direction. The left plate is stationary while the right plate moves at a constant speed $U_y$. The fluid is driven by the plate due to the viscous stress. This flow process is simulated by using the discrete model without considering the non-ideal gas effects and surface tension.

\begin{figure}
  \centering
  \includegraphics[width=0.9\textwidth]{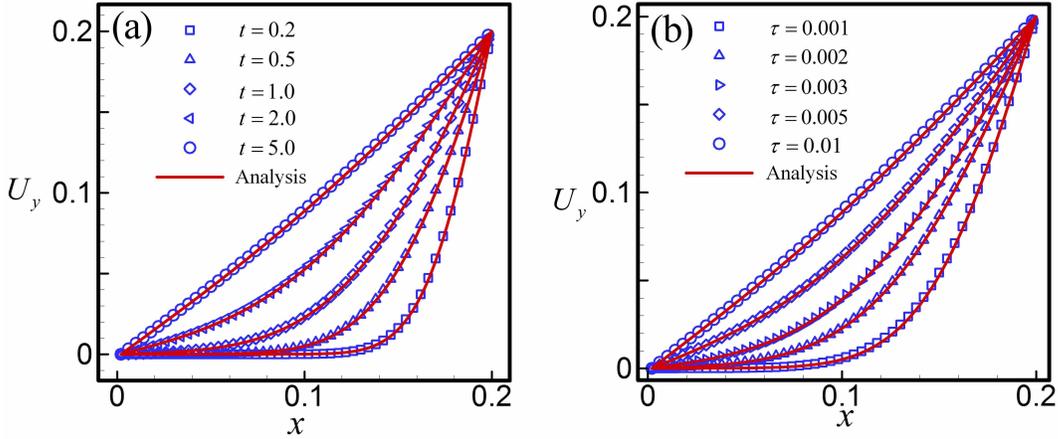}\\
  \caption{Verification of viscosity stress calculated by the new model. (a) Velocity profiles of Couette flow at several different times, (b) Velocity profiles of Couette flow with several different viscosity coefficients represented by relaxation time $\tau$. The symbols are DBM results while the solid lines are analytical solution.}\label{fig2}
\end{figure}

The speed of the moving plate is $U_y=0.2$ and the temperature of the whole flow field is fixed at $T=1.0$. The computational meshes are $N_x \times N_y=100 \times 1$ with spatial steps $dx=dy=0.002$ and time step $dt=0.0001$. Different viscosity coefficients are obtained by changing the relaxation time $\tau$. The initial velocity in the flow filed is zero. The non-slip boundary condition is used on both left and right boundary. The first-order forward difference is used for time discretization and the NND scheme is used for spatial discretization. The simulation results are shown in Fig. \ref{fig2}.

Figure \ref{fig2} (a) gives the profiles of velocity $U_y$ along the $x$ direction at several different times with a fixed relaxation time $\tau=0.002$, while figure \ref{fig2} (b) shows profiles of $U_y$ with several different relaxation time at $t=1.0$. The analytical solutions denoted by the solid lines are also plotted for comparison. The simulation results are in good agreement with the analytical solutions, which verifies the accuracy of the viscous stress calculated by the new discrete model.

\subsection{Two-phase coexistence curve}
\begin{figure}
  \centering
  \includegraphics[width=0.9\textwidth]{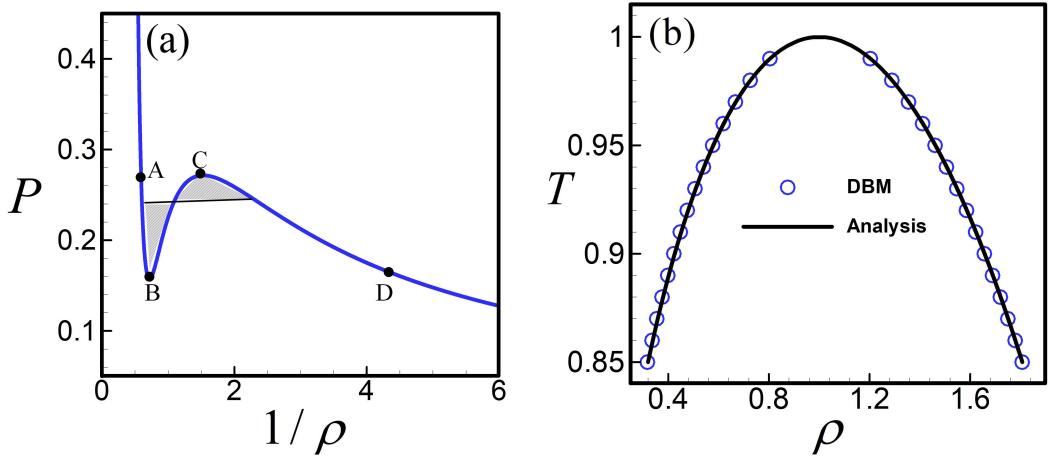}\\
  \caption{(a) $P$ versus $1/\rho$ for the equation of state of van der Waals, (b) Verification of non-ideal equation of state by two-phase coexistence curve.}\label{PV1}
\end{figure}

The two-phase coexistence problem is one of the most typical multiphase flow problems that can be described by a single-component multiphase flow model. According to the VDW EOS, below the critical temperature $T_c$, the system allows for two coexisting fluid with different densities under the same pressure $P$. The high density ($\rho_l$) corresponds to the liquid phase, while the lower density ($\rho_v$) to the vapor phase. Seen from the EOS shown in Fig. \ref{PV1} (a), there are many groups of coexistence points at different pressures, as long as the value of pressure is between $A$ and $B$; the liquid phase corresponds to the $AB$ segment while the vapor phase to the $CD$ segment. However, only one set of liquid-vapor points satisfy both the mechanical equilibrium and thermodynamic equilibrium (more specifically, the chemical potential equilibrium) at a given temperature $T_0$. This set of liquid-vapor points can be calculated by the Maxwell equal area rule.

The liquid-vapor coexistence points at several different temperatures are calculated by using the new multiphase flow model. The computation meshes are $N_x \times N_y=250 \times 1$ with the spatial steps $dx = dy = 4 \times 10^{-3}$. The time step and the relaxation time are $dt=5 \times 10^{-5}$ and $\tau =0.02$, respectively. The VDW EOS in Eq. (\ref{Eq-EosVDW}) is adopted with $a=9/8$ and $b=1/3$. The coefficient of surface tension is $K=2 \times 10^{-5}$. The first-order forward difference is used for time discretization. The NND scheme is adopted to calculate the spatial derivative in the convection term while the NPS scheme is used to calculate the spatial derivatives in the force term $\mathbf{F}'$. Periodic boundary conditions are adopted on both left and right boundaries. The initial conditions are set as follows:
\begin{equation}\label{Eq-test2}
\left\{ \begin{array}{l}
{\left( {\rho ,T,{u_x},{u_y}} \right)_L} = \left( {{\rho _v},0.99,0,0} \right)\\
{\left( {\rho ,T,{u_x},{u_y}} \right)_M} = \left( {{\rho _l},0.99,0,0} \right)\\
{\left( {\rho ,T,{u_x},{u_y}} \right)_R} = \left( {{\rho _v},0.99,0,0} \right)
\end{array} \right.
\end{equation}
where the subscript ``L'', ``M'', and ``R'' indicate the regions $0< x\leq N_x/4$, $N_x/4<x \leq 3N_x/4$, and $3N_x/4 < x \leq N_x$, respectively. $\rho_v$ and $\rho_l$ are the theoretical vapor and liquid densities at $T=0.99$. According to the VDW EOS combined with the equal area rule we get that $\rho_v=0.8045$ and $\rho_l=1.2035$.

The simulation continues until the equilibrium state is achieved, then the temperature drops by $0.01$ and wait for the system to achieve another equilibrium state. The above process is repeated until the temperature drops to $T=0.85$. The vapor and liquid densities calculated by the new model at different temperatures are plotted in Fig. \ref{PV1} (b). The symbols are simulated by the new multiphase flow model while the solid line represents the analytic solution. The simulation results are in excellent agreement with the theoretical solutions, which verified the accuracy of the new multiphase model in calculating the equilibrium phase transition.

\begin{figure}
  \centering
  \includegraphics[width=0.6\textwidth]{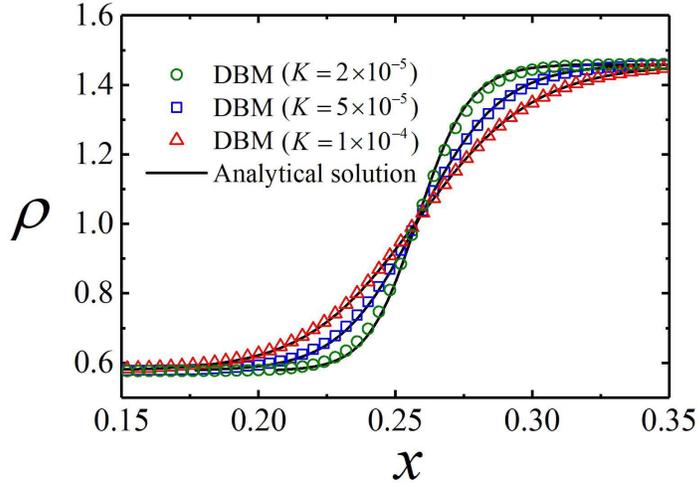}\\
  \caption{The density interface between the liquid and vapor phases at $T=0.95$. The symbols are calculated by DBM and the solid lines are theoretical solutions.}\label{density-profile}
\end{figure}

To further validate the effect of surface tension, the transition curves between the liquid and vapor phases with different coefficients of surface tension, i.e., $K=2 \times 10^{-5}$, $K=5 \times 10^{-5}$, and $K=1 \times 10^{-4}$, at $T=0.95$ are calculated and the results are shown in Fig. \ref{density-profile}. The theoretical solutions are also plotted for comparison. The theoretical solution of phase interfaces given by the modified VDW theory reads \cite{Bongiorno1975}
\begin{equation}\label{Eq-Analysisprofile1}
x - {x_0} =  - \frac{1}{{\sqrt {2a/K} }}\int_{{\rho ^*}({x_0})}^{{\rho ^*}(x)} {\frac{1}{{\sqrt {{\Phi ^*}({\rho ^*}) - {\Phi ^*}(\rho _l^*)} }}} d{\rho ^*}
\end{equation}
with
\begin{equation}\label{Eq-phixingrho}
{\Phi ^*}({\rho ^*}) = {\rho ^*}\xi  - {\rho ^*}{T^*}\left[ {\ln (1/{\rho ^*} - 1) + 1} \right] - {({\rho ^*})^2}
\end{equation}
and
\begin{equation}\label{Eq-phixingrho}
\xi  = {T^*}\ln (1/\rho _s^* - 1) - \frac{{\rho _s^*{T^*}}}{{1 - \rho _s^*}} + 2\rho _s^*
\end{equation}
where $\rho^*$ and $T^*$ denotes the reduced density and temperature, $\rho^*=b \rho$ and $T^*=b T/a$, $\rho _s^*$ represents the reduced density of the liquid or vapor corresponding to the equilibrium state at $T^*$.

The theoretical solutions are shown as solid lines and the DBM results are represented by symbols. It shows that the density profiles of the interface simulated by the new multiphase model agree well with the theoretical solutions, which verifies the accuracy of the surface tension of the new model.

\subsection{Droplet suspension}
To further validate the effect of surface tension, a circular droplet suspended in its vapor is simulated. According to the Laplace law, under a fixed surface tension coefficient, the pressure difference $\Delta P$ inside and outside the droplet is inversely proportional to the radius of the droplet $R_0$, i.e., $\Delta P \sim 1/R_0$.

\begin{figure}
  \centering
  \includegraphics[width=0.8\textwidth]{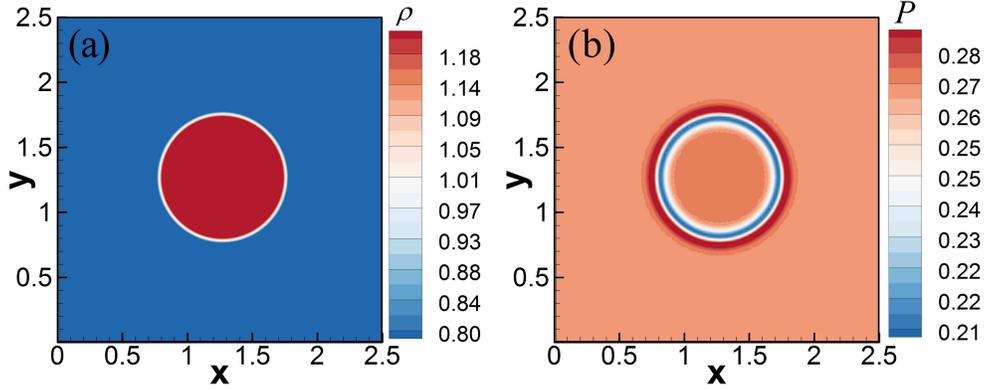}\\
  \caption{Simulation results of circular droplets suspended in its vapor: (a) density contour map and (b) pressure contour map.}\label{fig3-1}
\end{figure}

Several droplets with various radii are simulated under two different coefficients of surface tension $K=2\times 10^{-5}$ and $K=2\times 10^{-4}$, respectively. The computational meshes are $N_x \times N_y =128\times128$ with the spatial steps $dx=dy=0.02$ and time step $dt=0.00001$. Periodic boundary conditions are adopted on the boundaries and corners. The first-order forward difference is used for time discretization. The FFT scheme is used to calculate the space derivative in both the convection term and the force term. As a representative, the contour maps of density and pressure for a droplet with $R_0=0.5$ are shown in Fig. \ref{fig3-1}(a) and (b), respectively.

\begin{figure}
  \centering
  \includegraphics[width=0.8\textwidth]{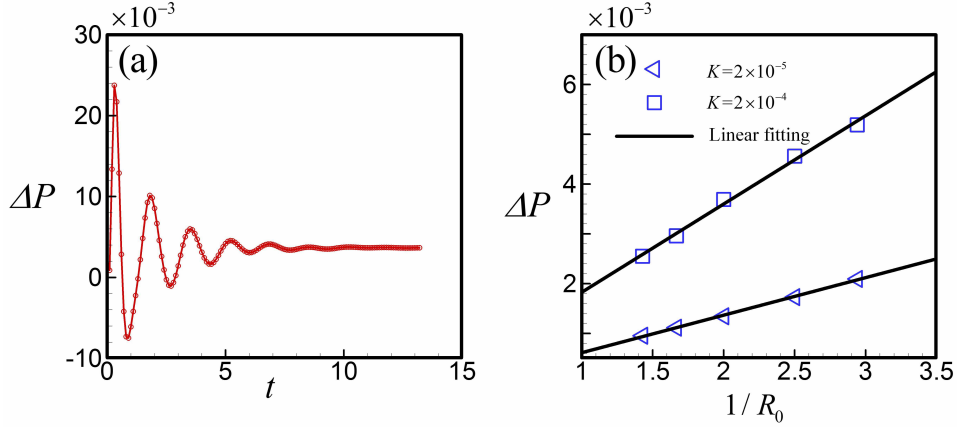}\\
  \caption{Verification of surface tension by the Laplace's law: (a) evolution of pressure difference $\Delta P$ between the inside and the outside the droplet with time $t$ and (b) the relationship between $\Delta P$ and the radius of droplet $R_0$.}\label{fig3-2}
\end{figure}

The pressure difference $\Delta P$ inside and outside the circular droplet is detected during the calculation. The profile of $\Delta P$ with time is shown in Fig. \ref{fig3-2}(a) and it eventually tends to a stable value. The calculation continues until $\Delta P$ is almost unchanged, then the pressure difference $\Delta P$ and the corresponding radius $R_0$ is recorded. Several circular droplets with different radii are simulated. Figure \ref{fig3-2}(b) shows the relation between the $\Delta P$ and $1/R_0$, in which symbols are simulation results and the solid lines are linear fitting. Clearly, $\Delta P$ is proportional to $1/R_0$, which agrees well with the Laplace law. The accuracy of the surface tension calculated by the new multiphase model is well verified from this simulation.

\subsection{Phase separation}
\begin{figure}
  \centering
  \includegraphics[width=0.8\textwidth]{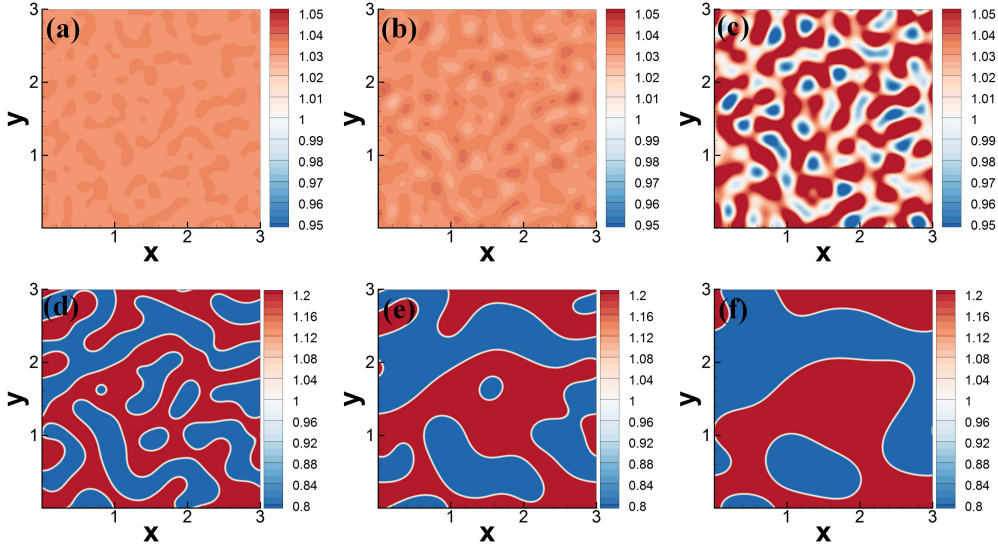}\\
  \caption{Density contour maps in the process of isothermal phase separation at (a) $t=0.5$, (b) $t=1.0$, (c) $t=2.0$, (d) $t=5.0$, (e) $t=9.0$, and (f) $t=14.0$, respectively.}\label{Phase-seperation1}
\end{figure}

In this part, a dynamic phase separation process is simulated by using the new multiphase flow model. The initial conditions are set as
\begin{equation}\label{Eq-test4}
({\rho ,T,{u_x},{u_y}})=(1.0+\delta, 0.92, 0, 0)
\end{equation}
where $\delta$ is a random noise with amplitude 0.01, which provides a starting point for phase separation. In the phase separation process, the temperature is fixed at $T=0.92$, which is equivalent to that the system contacts with a large heat source with a temperature $T=0.92$. The computational meshes are $N_x \times N_y = 300 \times 300$ with spatial steps $\Delta x=\Delta y=0.01$. The time step and relaxation time are $\Delta t= 1\times 10^{-4}$ and $\tau=0.01$, respectively. Periodic boundary conditions are adopted on all boundaries and corners. The first-order forward difference is used for time discretization. The NND and NPS schemes are used to calculate the spatial derivatives in the convection term and the force term. Figure \ref{Phase-seperation1} shows the density contour maps at several different times. The simulation results are consistent with those in previous literature\cite{free-energy1995,GanSoftmatter2016,SoftMatter2019}.

\subsection{Droplets collision}
Droplet collision is another typical multiphase flow problem and has a broad application background in industrial production. It involves complex interfacial dynamics, including interface movement, fusion, and separation, etc. In this part, we simulate the head-on collision between two droplets using the new multiphase flow model. Two droplets with the opposite speeds are suspended in the same horizontal position. The droplet on the left moves to the right at speed $U_0$, while the droplet on the right moves to the left at the same speed. There are two situations after the collision depending on the initial speed $U_0$. If $U_0$ is much small, the two droplets eventually fuse together, whereas if $U_0$ is large enough, they separate again along the vertical direction.

\begin{figure}
  \centering
  \includegraphics[width=0.8\textwidth]{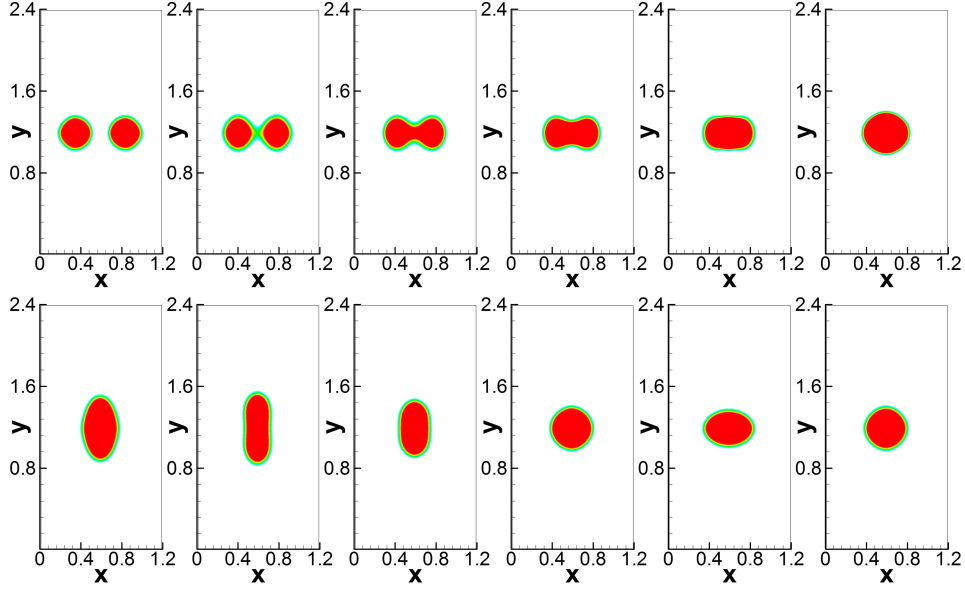}\\
  \caption{Head-on collision of two droplets which are fused together after the collision.}\label{fig4-2}
\end{figure}

\begin{figure}
  \centering
  \includegraphics[width=0.8\textwidth]{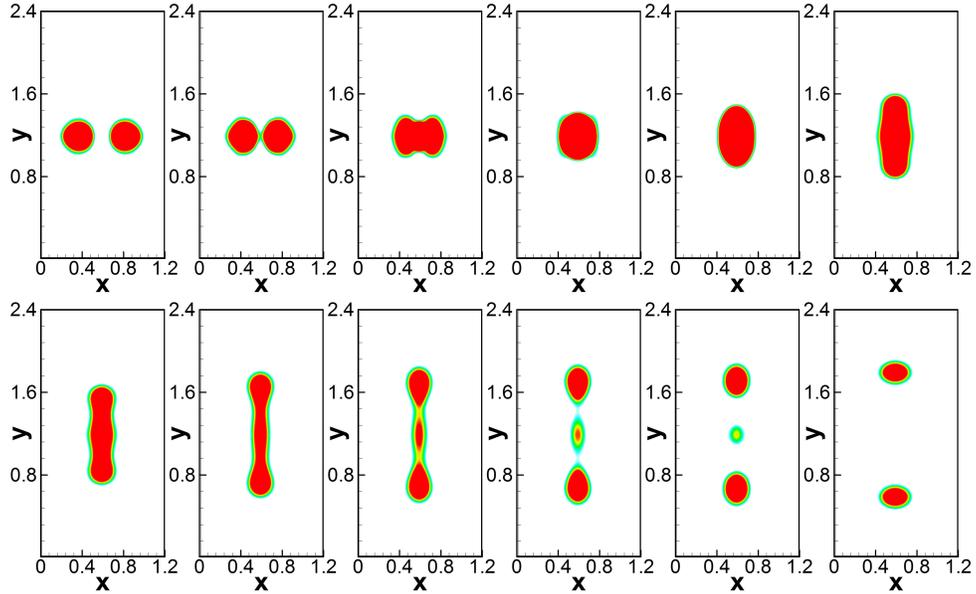}\\
  \caption{Head-on collision of two droplets which are separated after the collision.}\label{fig4-3}
\end{figure}

Two cases with $U_0=0.2$ and $U_0=0.5$, respectively, are simulated by using the new multiphase flow model. The computational meshes are $N_x \times N_y =60 \times 120 $ with the spatial step $\Delta x=\Delta y=0.02$. The time step is $\Delta t=1 \times 10^{-4}$ and the relaxation time is $\tau=2 \times 10^{-3}$. The coefficient of surface tension is set as $K=2 \times 10^{-4}$ and the temperature is fixed at $T=0.92$. The boundary conditions and the discrete schemes are the same as the previous simulation of phase separation. The snapshots of the collision processes for two cases are shown in Figs. \ref{fig4-2} and \ref{fig4-3}, respectively. Figure \ref{fig4-2} shows the case with $U_0=0.2$, in which two droplets fuse together after the collision and figure \ref{fig4-3} shows the case with $U_0=0.5$ where they separate after the collision. The interfacial dynamics in droplet collision, fusion, and separation are well described, and the simulation results are consistent with the previous studies \cite{Huang-book}.

\begin{figure}
  \centering
  \includegraphics[width=0.9\textwidth]{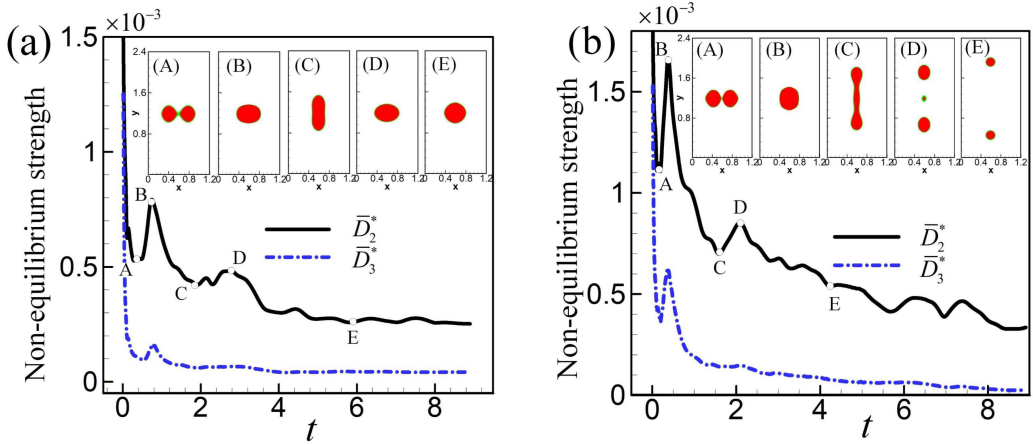}\\
  \caption{Profiles of two kinds of non-equilibrium strength in the collision process, where (a) is for head-on collision of two droplets which are fused together after the collision,
  (b) is for head-on collision of two droplets which are separated after the collision.
  It is clear that $\bar{D}_2$ can be used to characterize the stage behaviors of the collision process and $\bar{D}_2$ shows significantly different behaviors in the last stage ( after the ``D" point in the $\bar{D}_2^* (t)$ curve ).
  }\label{fig-Detltaxing2-3}
\end{figure}

The evolutionary characteristics of non-equilibrium strength quantities, $D^*_2$ and $D^*_3$, are shown in Fig. \ref{fig-Detltaxing2-3}, in which the left and right subgraphs correspond to the two kinds of droplets collision with $U_0=0.2$ and $U_0=0.5$, respectively. The spatial mean values $\bar{D}^*_2$ and $\bar{D}^*_3$ are used to represent the total mean non-equilibrium strength, i.e.
\begin{equation}\label{Eq-Dxingbar}
\bar D_m^{\text{*}} = \frac{1}{{{l_x}{l_y}}}\iint {D_m^{\text{*}}dxdy},  {\kern 26pt} m = 2,3
\end{equation}
where $l_x l_y$ represents the total area of the simulation region.

Figure \ref{fig-Detltaxing2-3} (a) shows that at least five stages of the collision can be identified from the non-equilibrium strength $\bar{D}^*_2$. In the first stage, two droplets come close to each other, and $\bar{D}^*_2$ gradually decreases until it reaches point ``A'' in the $\bar{D}_2^* (t)$ curve. In the second stage (AB segment on the profile of $\bar{D}^*_2$), two droplets contact and the interface begins to fuse, and $\bar{D}^*_2$ increases with time. After the phase interface fusion is completed, the third stage (BC segment) begins, and the merged droplet elongate vertically, accompanied by a decrease of $\bar{D}^*_2$. Because the kinetic energy after collision is not enough to overcome the constraint of surface tension, the merged droplet can not be separated again. As a result, in the fourth stage (CD segment), the droplet contracts vertically and the overall trend of the non-equilibrium strength is upward. In the fifth stage (after the ``D" point in the $\bar{D}_2^* (t)$ curve), the droplet lengthens and contracts alternately in the horizontal and vertical directions, and finally tends to a stable state. Correspondingly, the $\bar{D}^*_2$ shows the trend of oscillation attenuation and finally tends to a stable value. From the profile of $\bar{D}^*_3$, only the first three stages can be well identified, which indicates that the roles of non-equilibrium strength of different orders are much different.

For the second type of droplet collision with $U_0=0.5$, there are also five stages as shown in Fig. \ref{fig-Detltaxing2-3} (b). The profile of $\bar{D}^*_2$ in the first four stages is similar to the case with $U_0=0.2$. However, in the last stage,  $\bar{D}^*_2$ continues to decrease until the separated droplets leave the simulation region at $t=5$. The profiles after $t=5$ is meaningless because the droplet has left the simulation region. So these two types of collisions can be well distinguished by the last stage of $\bar{D}^*_2$. Likewise, from the profile of $\bar{D}^*_3$, only the first three stages can be well identified. As a result, it is difficult to distinguish these two types of collision cases from the profile of $\bar{D}^*_3$.

\begin{figure}
  \centering
  \includegraphics[width=0.7\textwidth]{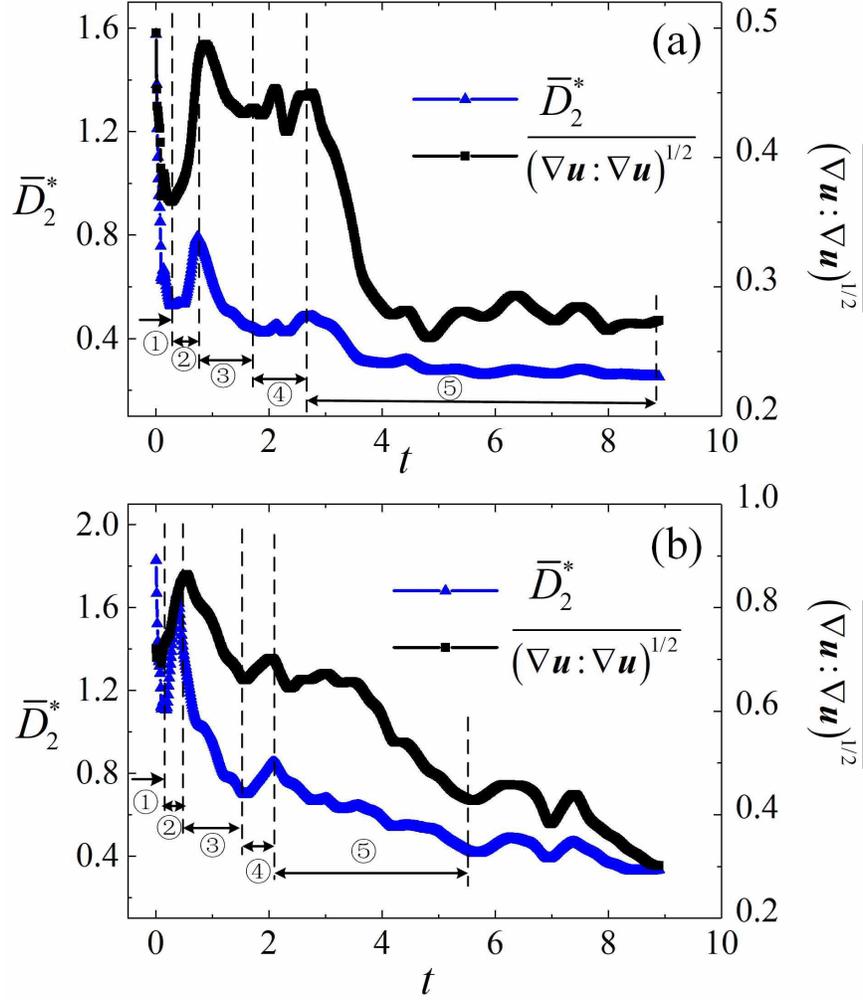}\\
  \caption{Profiles of $\bar{D}_2^*$ and $\overline {{{\left( {\nabla {\bf{u}}:\nabla {\bf{u}}} \right)}^{1/2}}} $ in the collision process, where (a) is for head-on collision of two droplets which are fused together after the collision, (b) is for head-on collision of two droplets which are separated after the collision.
  }\label{fig-D2xing-Deltau}
\end{figure}

The evolution of the non-equilibrium strength with time can be explained by the change of the velocity gradient in the flow field. Because ${\bf{\Delta }}_2^*$ corresponds to the viscous stress ${\bf{\Pi }}$ in the hydrodynamic equations, which can be expressed as Eq. (\ref{Eq-viscousstress}) in the NS level. In this approximation, $D_2^*$ can be expressed as
\begin{equation}\label{Eq-D2xing-NS}
D_2^* \approx {(2\mu )^{1/2}}{\left[ {\nabla {\bf{u}}:\nabla {\bf{u}} + 2(\frac{{\partial {u_x}}}{{\partial y}}\frac{{\partial {u_y}}}{{\partial x}} - \frac{{\partial {u_x}}}{{\partial x}}\frac{{\partial {u_y}}}{{\partial y}})} \right]^{1/2}}.
\end{equation}
We further find that the last two terms on Eq. (\ref{Eq-D2xing-NS}) have little effect on the overall evolutionary characteristics. For the sake of convenience, we will only use ${\left( {\nabla {\bf{u}}:\nabla {\bf{u}}} \right)^{1/2}}$ in the following analysis. Figure \ref{fig-D2xing-Deltau} shows the profiles of $\bar{D}_2^*$ and the spatial average of ${\left( {\nabla {\bf{u}}:\nabla {\bf{u}}} \right)^{1/2}}$. The definition of $\overline {{{\left( {\nabla {\bf{u}}:\nabla {\bf{u}}} \right)}^{1/2}}} $ is the same as that of $\bar{D}_m^*$ in Eq. (\ref{Eq-Dxingbar}). Figures \ref{fig-D2xing-Deltau} (a) and (b) correspond to the first and second types of collision processes, respectively. It shows that $\bar{D}_2^*$ and $\overline {{{\left( {\nabla {\bf{u}}:\nabla {\bf{u}}} \right)}^{1/2}}} $ have similar evolutionary characteristics in both Fig. \ref{fig-D2xing-Deltau} (a) and (b), which is consistent with our previous analysis.

For the first type of droplet collision, the collision process is roughly divided into five stages as shown in  Fig. \ref{fig-D2xing-Deltau} (a). In the first stage, the two droplets approach each other independently. Part of the kinetic energy of the droplets is transferred to the flow field of vapor phase, leading to a decrease of the system's velocity gradient whole. In the second stage, two droplets collide and fuse. During the fusion process, the interface contracts rapidly due to the work done by surface tension, which results in the increase of velocity gradient in the flow field. After the droplets are completely fused, the new droplet begins to stretch in the $y$ direction in the third stage. In this process, the surface tension does the negative work, causing the interface to slow down and the velocity gradient to decrease. In the fourth stage, because the kinetic energy of the interface motion is not enough to overcome the constraint of surface tension, the interface begins to contract in reverse when it stretches to the longest in the $y$ direction. The surface tension does positive work, so the interface accelerates and the velocity gradient increases. Soon the interface shrinks to its minimum in the $y$ direction and begins to stretch again in the $x$ direction. The velocity gradient decreases in the process of stretching and increases in the process of shrinking. Finally, in the fifth stage, the droplet is alternately stretched and contracted in the $x$ and $y$ direction, but the maximum position it can reach gradually shortens until it reaches a stable state.

For the second type of droplet collision, the first stages are the same as those of the first type of collision. However, in the fourth stage, the fused droplet rupture and separate again. The interface of the two separated droplets shrinks rapidly, leading to an increase of velocity gradient. After that, in the fifth stage, the single droplet decelerates independently, so the velocity gradient decreases until the droplets leave the simulation area. ${\bf{\Delta }}_3^*$ is a higher-order non-equilibrium characteristic quantity related to heat flux but its corresponding macroscopic physical significance is not yet clear. Because the temperature is set as a constant in the simulation, the heat flow in the flow field is sharply limited, so the amplitude is relatively small and its evolution characteristics are not visible.

\section{Conclusion}\label{sec4}
In this work, under the framework of the discrete Boltzmann method, a new kinetic multiphase flow model based on a discrete Enskog equation was proposed. Compared with the previous multiphase flow discrete Boltzmann model, a bottom-up modeling approach was adopted in the new model. The repulsion potential was introduced by using the Enskog collision operator instead of the Boltzmann collision operator, and the attraction potential is incorporated through average filed approximation. The effect of total intermolecular forces is ultimately taken into account via the force term of the DBM.
The discrete local equilibrium distribution function is solved by the Hermite polynomials. The third order Hermite polynomial is adopted in order to improve the computational efficiency as much as possible, although it is straightforward to take the higher-order ones. Several benchmarks, including the Couette flow, two-phase coexistence curve, droplet suspension, phase separation, and droplet collisions are simulated, through which the accuracy of the new multiphase flow model is verified. In addition, for two kinds of droplet collisions, the non-equilibrium characteristics are comparatively investigated via two TNE strength quantities, $\bar{D}_2^*$ and $\bar{D}_3^*$. It is found that during the collision process, $\bar{D}_2^*$ is always significantly larger than $\bar{D}_3^*$, $\bar{D}_2^*$ can be used to identify the different stages of the collision process and to distinguish different types of collisions.


%
%

\section*{Acknowledgements}
Many thanks to the anonymous referees for their helpful comments and suggestion. This work is supported by the China Postdoctoral Science Foundation under Grant No. 2019M662521, National Natural Science Foundation of China under Grant No. 11772064, CAEP Foundation (under Grant No. CX2019033) and the opening project of State Key
Laboratory of Explosion Science and Technology (Beijing Institute
of Technology) under Grant No. KFJJ19-01 M.
%
%

\end{document}